\documentclass[9pt,journal]{IEEEtran}
\usepackage{lineno,hyperref}
\usepackage{amssymb}
\usepackage{graphicx}
\usepackage[utf8]{inputenc}
\usepackage[T1]{fontenc}

\usepackage{amsmath}
\usepackage
[
    ,letterpaper
    ,left       = 1in
    ,right      = 1in
    ,top        = 1.5in
    ,headheight = 1in
    ,headsep    = .3in
]{geometry}

\modulolinenumbers[5]

\begin{document}

\title{Capacity Enhancement of Cooperative NOMA over Rician Fading Channels with Orbital Angular Momentum 
{\footnotesize \textsuperscript{}}
\thanks{}
}


\author{Ahmed Al Amin, \IEEEmembership{}
Bhaskara Narottama
       and\IEEEmembership{}
       Soo Young Shin,~\IEEEmembership{Senior Member,~IEEE}
\IEEEcompsocitemizethanks{\IEEEcompsocthanksitem -This work was supported by Priority Research Centers
Program through the National Research Foundation of Korea (NRF) funded by the Ministry of Education, Science and Technology (2018R1A6A1A03024003). 
\IEEEcompsocthanksitem -The authors are with the WENS Laboratory, Department of IT Convergence Engineering, Kumoh National Institute of Technology, Gumi 39177, South Korea (e-mail: amin@kumoh.ac.kr; bhaskaranarottama@gmail.com; wdragon@kumoh.ac.kr).\protect\\

 }
\thanks{ 
}}

\maketitle

\begin{abstract}
This letter proposes the usage of orbital angular momentum (OAM) for cooperative non-orthogonal multiple access (CNOMA) to enhance sum capacity (SC) for the future cellular communication system. 
The proposed CNOMA-OAM scheme is analyzed and compared with other schemes, i.e., conventional CNOMA, conventional orthogonal multiple access (OMA) with OAM. 
The impact of the power allocation factor for OAM beam over SC is also analyzed. The analytical result is justified by simulation results which demonstrate that the proposed CNOMA-OAM  provides higher SC compared to other schemes.
\end{abstract}
\begin{IEEEkeywords}
Orbital Angular Momentum (OAM), cooperative non-orthogonal multiple access (CNOMA), orthogonal multiple access (OMA), sum capacity.
\end{IEEEkeywords}
\section{Introduction}
Channel capacity becomes an important criterion as the next generation of wireless communications are expected to handle a 1000-fold increase in data traffic [1]. Power domain non orthogonal multiple access (NOMA) provides higher capacity gain than other conventional multiple access methods. Power-domain NOMA allows different information signals to be super-positioned and transmitted to several users symbol simultaneously.
Information signals for the cell center user (CCU) and the cell edge user (CEU) are differentiated by power allocation level [2].
Cooperative NOMA (CNOMA) enhances data reliability and coverage area by utilizing CCU as a relay for CEU. CCU decodes its own signal by SIC. It also decodes the symbol of CEU and relaying to the CEU to enhance reliability [3].   
There have been works on CNOMA to improve the channel capacity. Previous works proposed  
the use of user pairing [4], the combination of orthogonal multiple access (OMA) and NOMA [5],
integration of generalized space shift keying (GSSK) with NOMA [6],
and the use of coordinated multiple points (CoMP) and NOMA [7].
There is a huge potential to utilize orbital angular momentum (OAM) signal to improve the sum capacity (SC) of CNOMA. 
OAM utilizes a new degree of freedom which is known as OAM mode for signal transmission [8-9]. OAM exploits the phase variation with respect to the azimuth angle of the propagated electromagnetic waves. This leads to the helical phase structure of the wave. 
The considered interplay between CNOMA and OAM is named as CNOMA-OAM in this letter.   
 The contribution of this letter can be summarized as follows.
\begin{itemize}
\item A CNOMA-OAM is proposed to enhance the SC. 
\item Mathematical analysis is presented to inspect the SC of CNOMA-OAM scheme.
\item The impact of allocated power of OAM beam over the SC is analyzed.
\item 
Performance comparison in terms of SC for CNOMA-OAM method against conventional CNOMA and OMA-OAM over Rician fading channel is also analyzed.
\end{itemize}
The rest of this letter is organized as follows. Section II describes the system model and the proposed CNOMA-OAM. The other schemes are also presented in this section. 
Section III mathematically analyzes the SC of the proposed CNOMA-OAM. 
Section IV exhibits simulation results. 
This paper is concluded in Section V. 
\section{System and Protocol Descriptions}
\begin{figure}[t!]
\centering
\includegraphics[width=0.4\textwidth]{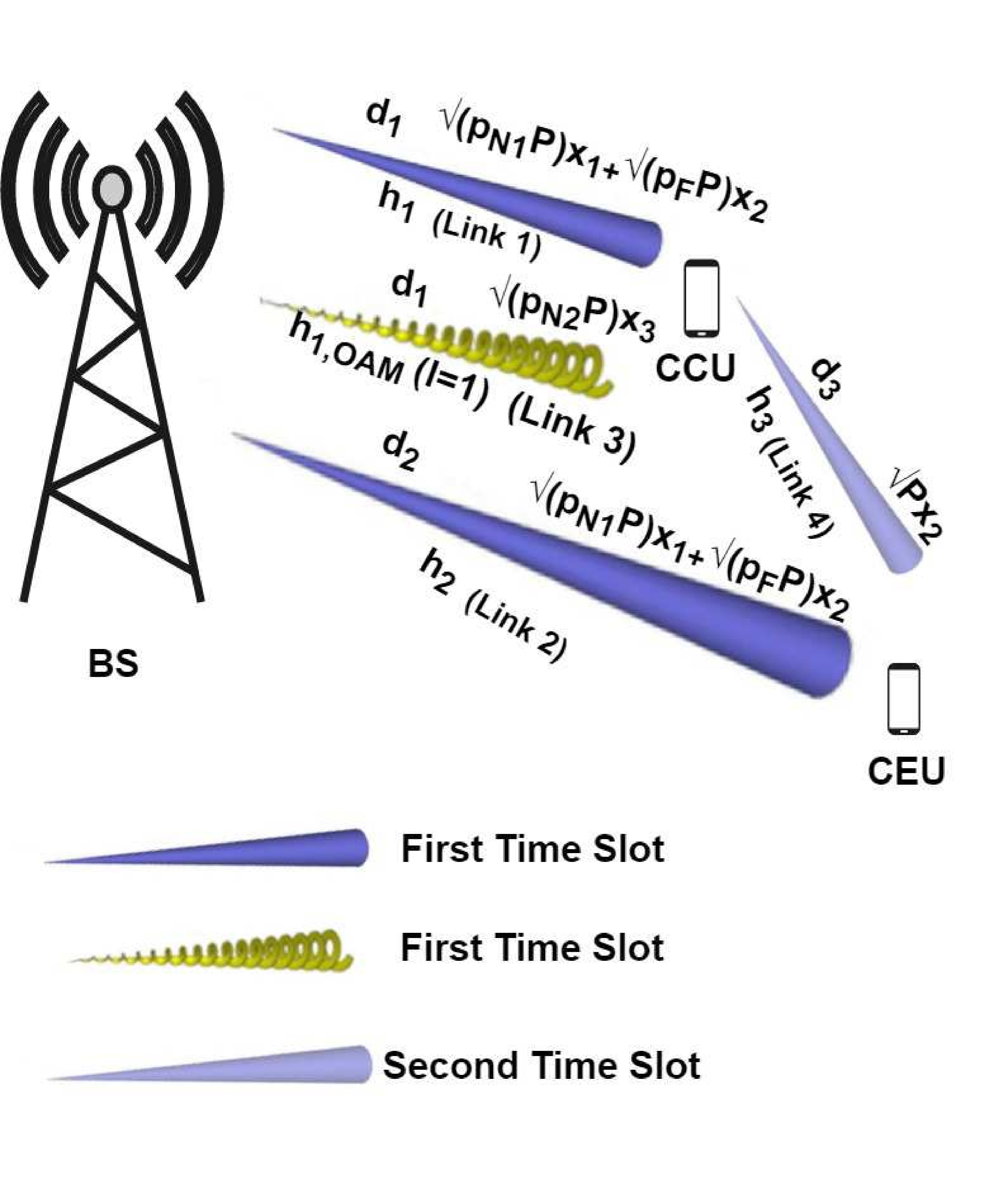}
\caption{System model of CNOMA-OAM scheme.}
\label{image-myimage}
\end{figure}
The proposed CNOMA-OAM for down-link (DL) communication is presented in Fig. 1. 
A base station (BS) and two users, i.e., a cell center user (CCU) and a cell edge user (CEU), in a single cell is assumed. 
CCU is assumed to be located much closer to the BS than CEU. 
Perfect channel state information (CSI) is assumed. 
CCU is assumed to have higher channel gain compared to CEU.
Moreover, to enhance data reliability and coverage area, CCU is used as decode and forward (DF) relay for CEU [8]. 
$d_1$ and $d_2$ are the respective normalized distances of CCU and CEU from BS.
Independent Rician fading channel coefficients of BS-to-CCU (Link 1), BS-to-CEU (Link 2), and CCU-to-CEU (Link 4) are denoted as $h_1$, $h_2$, and $h_3$, respectively. 
$\Omega_{CCU}$, $\Omega_{CEU}$, and $\Omega_{Relay}$ are the average power of BS-to-CCU, BS-to-CEU, and CCU-to-CEU link, respectively [8]. 
$\Omega_{CEU}<\Omega_{CCU}$ is assumed here. 
Moreover, the OAM channel from the BS-to-CCU (Link 3) is denoted as $h_{1, OAM}$.
Moreover, $l$ represents the considered OAM mode.
To reduce attenuation, low number of OAM mode, i.e. $l=1$, is considered [11].
Since OAM has better performance in line-of-sight (LOS) condition and the user with better channel condition, that is why OAM is only considered for CCU in Fig.1 [12].
$p_{N_1}$, $p_{N_2}$ and $p_F$ are the allocated powers for the CCU, OAM for CCU and CEU respectively. Where, $p_F$ \textgreater $p_{N_1}$+$p_{N_2}$ and $p_{N_1}=p_{N_2}=\frac{P-p_F}{2}$. $P$ is denoted as total transmit power [8].  
The proposed DL transmission protocol for CNOMA-OAM scheme is illustrated in Fig. 2. Where $T$ is the total duration of a time slot for total DL transmission. According to the proposed protocol, $x_1$ and $x_2$ is transmitted by CNOMA to CCU and CEU respectively in first time slot of T/2 duration. Moreover, at the same time slot $x_3$ is transmitted simultaneously to CCU by OAM without any interference. In the second time slot of T/2 duration, CCU (DF) relaying $x_2$ to CEU to enhance the reliability.  
\subsection{Direct Transmission}
According to the CNOMA concept, in first time slot BS transmits the superposition of two different data symbols $x_1$ and $x_2$ to the CCU and CEU as below
\begin{equation}
A=\sqrt{p_{N_1}P}x_1+\sqrt{p_{F}P}x_2,
\end{equation}
where $x_i$ denotes the i-th data symbol [8]. The CCU acquires $x_1$ from Eq. (1) by using SIC. 
The received signal-to-interference plus noise ratio (SINR) for symbols $x_1$ and $x_2$ at CCU can be achieved as
\begin{equation}
\gamma_\text{CCU}^{1}={\rho{{|h_1|}^2}p_{N_1}},
\end{equation}
\begin{equation}
\gamma_\text{CCU}^{2}=\frac{\rho{{|h_1|}^2}p_F}{\rho{{|h_1|}^2}p_{N_1}+1},
\end{equation}
where, $\rho \triangleq \frac{P}{\sigma^2}$ is the transmit signal-to-noise ratio (SNR) and additive white Gaussian noise (AWGN) noise variance is $\sigma^2$ for all the received signal in this letter [8]. Moreover, $x_2$ is directly decoded by the CEU. The received SINR for symbol $x_2$ at CEU is obtained as [8]
\begin{equation}
\gamma_\text{CEU}^2=\frac{\rho{{|h_2|}^2}p_F}{\rho{{|h_2|}^2}p_{N_1}+1}.
\end{equation}
\par
Furthermore, $x_3$ is directly transmitted using OAM ($l=1$) with $p_{N_2}$ to CCU simultaneously. 
Due to the assumed close distance between CCU and BS,  Rician fading channel is considered [8].
The NOMA link will not interfere with the OAM link due to the difference in OAM mode [13-15]. 
The received SINR for symbol $x_3$ at CCU can be expressed as 
\begin{equation}
\gamma_\text{CCU}^{3}={p_{N_2} \rho }{{\mu_k^2}}.
\end{equation}
Where, $\mu_k$ is the singular value of the OAM channel in decreasing order. OAM beam has divergence in its high-intensity region which caused attenuation [15-17]. 
By using Fresnel-zone-plate lenses antenna at BS, this issue can be mitigated without affecting the helical phase profile of OAM beam [15-17].
The normalized channel matrix $h_{1, OAM}$ can be expressed as follows [16] 
\begin{equation}
h_\text{1,OAM} = \frac{e^{-j 2 \pi (m-1) l}}{M}.
\end{equation}
where $m$ is the index of the receiving antenna and $M$ is the total number of receiving antennas [16]. 
Since one OAM beam is considered here to transmit $x_3$ by OAM, so there are no possibilities of intersymbol interference or intermodal interference among the OAM beams for LOS system [16]. 
\subsection{Relay Transmission}
In the second time slot, Link 4 is transmitted with total transmit power.
It is assumed that CCU can perfectly decode symbol $x_2$ in the first time slot [19]. 
The received SINR at CEU for symbol $x_2$ by relaying from the CCU can be expressed as 
\begin{equation}
\gamma_\text{CEU}^\text{Relay}={\rho{{|h_3|}^2}P}.
\end{equation}
\begin{figure}[!t]
\centering
\includegraphics[width=0.4\textwidth]{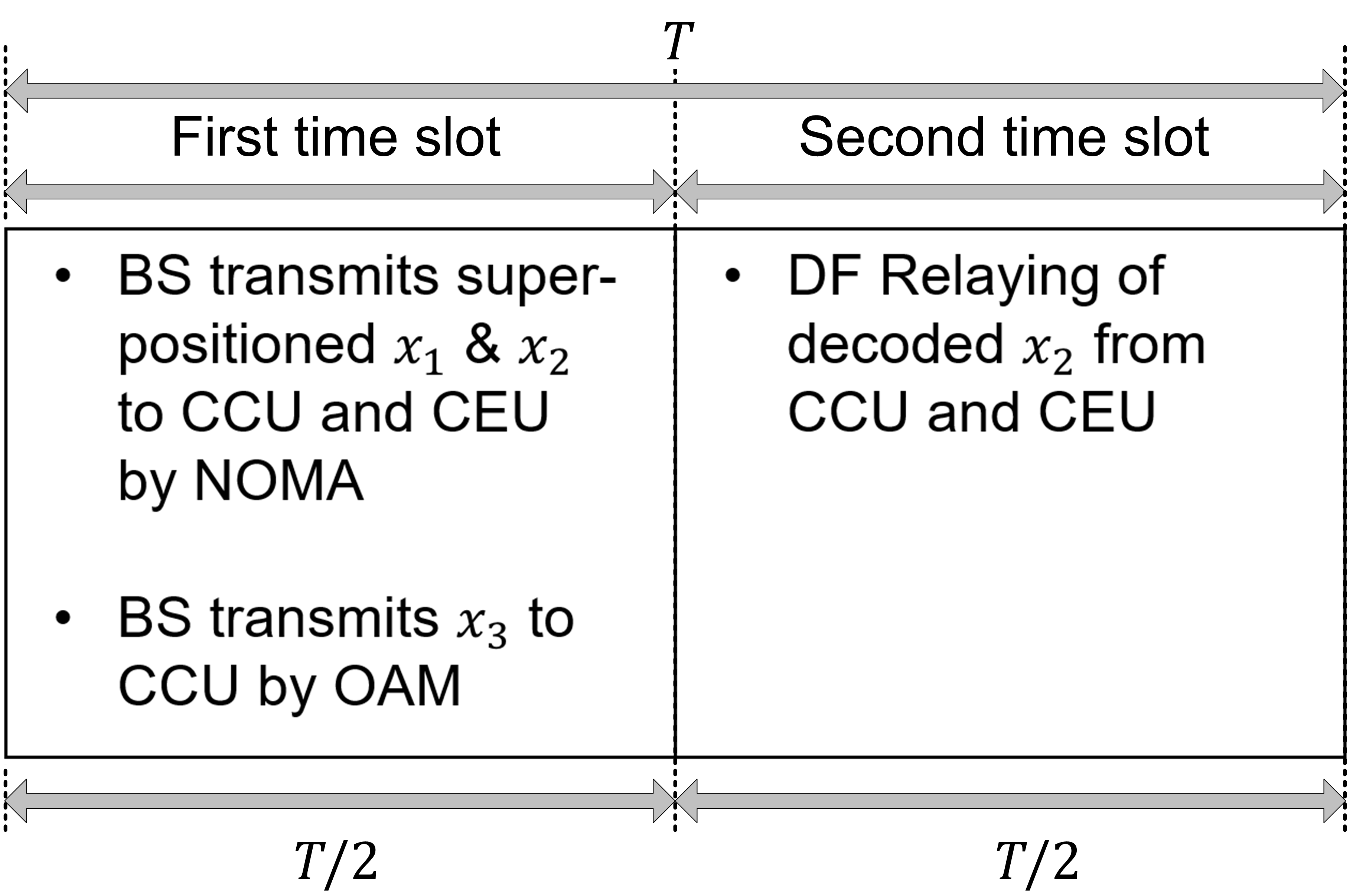}
\caption{Protocol for the proposed CNOMA-OAM.}
\label{image-myimage}
\end{figure}
\par
\subsection{Sum Capacity}
By considering normalized time and total transmit power ($T=1$ and $P=1$, respectively), the achievable capacity of $x_1$ and $x_2$ can be obtained as [8]
\begin{equation}
\begin{split}
C_{x_1} & =\frac{1}{2}\log_2(1+\min(|h_2|^2,|h_1|^2)\rho)\\
&-\frac{1}{2}\log_2(1+\min(|h_2|^2,|h_1|^2) p_{N_1} \rho ),
\end{split}
\end{equation}
\begin{equation}
C_{x_2}=\frac{1}{2}\log_2(1+\min(p_{N_1}|h_1|^2,|h_3|^2)\rho) ).
\end{equation}
Furthermore, the capacity of $x_3$ by OAM beam can be acquired as below
\begin{equation}
C_{x_3}=\frac{1}{2} \log_2(1+{p_{N_2} \rho }{{\mu_k^2}} ).
\end{equation}
where, K is the rank of the OAM channel matrix [16,21]. So, the sum capacity is given by
\begin{equation}
C_{CCU} = \text{E}[C_{x_1}]+\text{E}[C_{x_3}],
\end{equation}
\begin{equation}
C_{CEU} =\text{E}[C_{x_2}],
\end{equation}
\begin{equation}
C_\text{sum} = C_{CCU}+C_{CEU},
\end{equation}
where, E[.] represents the expectation operator. Whereas, $C_{CCU}$ and $C_{CEU}$ are the capacity of CCU and CEU for the proposed CNOMA-OAM scheme. The CNOMA-OAM scheme is compared with conventional CNOMA and OMA-OAM scheme as well over Rician fading channel. So the OMA-OAM is scheme is described in following subsection. 
\subsection{OMA-OAM Scheme}
For the OMA case, time division multiple access (TDMA) is considered.
In this case, BS transmits information signal for CCU and CEU independently in different time slots with total transmit power $P$. The different time slots allocated for CCU and CEU for different symbols (e.g. $x_1$, $x_2$ and $x_3$) and relay for $x_2$ are denoted as $t_1$, $t_2$, $t_3$ and $t_4$ respectively. Moreover, since $T=1$ is considered here so total time slot is equally splitted for each time slot for OMA-OAM scheme. Hence, $t_1=t_2=t_3=t_4=\frac{1}{4}$ is considered here. So, the achievable capacity of $x_1$ and $x_2$ can be achieved as below for the OMA-OAM scheme [8,19-20]
\begin{equation}
\begin{split}
C_{x_2}^\text{OMA}  = \frac{1}{4}\log_2(1+\min(|h_2|^2,|h_1|^2) P \rho ),
\end{split}
\end{equation}
\begin{equation}
C_{x_2}^\text{OMA}=\frac{1}{4}\log_2(1+\min(P |h_1|^2,|h_3|^2)\rho) ),
\end{equation}
Furthermore, the achievable capacity of $x_3$ by OAM beam can be acquired as below for OMA-OAM scheme [16-20]
\begin{equation}
C_{x_3}^{OMA}=\frac{1}{4}\log_2(1+{P \rho }{{\mu_k^2}} ).
\end{equation}
So, the SC of OMA-OAM scheme can be achieved by the following equation [8,19-20]
\begin{equation}
C_\text{CCU}^\text{OMA} = \text{E}[C_{x_1}^\text{OMA}]+\text{E}[C_{x_3}^\text{OMA}],
\end{equation}
\begin{equation}
C_\text{CEU}^\text{OMA} =\text{E}[C_{x_2}^\text{OMA}],
\end{equation}
\begin{equation}
C_\text{sum}^\text{OMA} =C_\text{CCU}^\text{OMA}+C_\text{CEU}^\text{OMA} ,
\end{equation}
where E[.] represents the expectation operator. Moreover, $C_\text{CCU}^\text{OMA}$ and $C_\text{CEU}^\text{OMA}$ are the capacity of CCU and CEU for OMA-OAM scheme. 
\section{Capacity Analysis}
In this section, closed form solution of the sum capacity of the proposed CNOMA-OAM over independent Rician fading channels is presented. Let $z_1 \triangleq \min(|h_1|^2,|h_2|^2)$ and $z_2 \triangleq \min(p_{N_1} |h_1|^2,|h_3|^2)$ [10]. So, the CDF of $z_1$ as 
\begin{multline}
F(z_1)=1-A_x A_y \sum_{k=0}^{\infty}\sum_{n=0}^{\infty} \tilde{B}_x(n) \tilde{B}_y(k)\\
\Gamma(n+1,a_x z_1)\Gamma(k+1,a_yz_1)
\end{multline}
where $B_x(n)=(K_x^n(1+K_x)^n)/(\Omega_x^n(n!)^2)$, 
$B_y(k)=(K_y^n(1+K_y)^k)/(\Omega_y^k(k!)^2)$, 
$a_x=(1+K_x)/\omega_x$, $a_y=(1+K_y)/\Omega_y$, $\tilde{B}_x(n)= B_x(n)/a_x^{n+1}$ and 
$\tilde{B}_y(k)= B_y(k)/a_y^{k+1}$. The subscript $x$ denotes BS to CEU link, $y$ denotes BS to CCU link and $w$ denotes CCU to CEU link respectively [8]. 
Similarly, the CDF of $z_2$ can be obtained as below from [8]
\begin{multline}
G(z_2)= 1- A_w A_y \sum_{n=0}^{\infty}\sum_{k=0}^{\infty} \tilde{B}_w(n) \tilde{B}_y(k) \\
 \Gamma (n+1,a_w z_2) \Gamma (k+1,a_y/p_{N_1} z_2)
\end{multline}
where parameters are defined as Eq. (16) which is expressed as before. So, the exact expression of $C_{x_1}$ can be obtained as below [10],
\begin{equation}
C_{x_1}^{exact} = \frac {1}{2ln2}(D(\rho)-D( p_{N_1}\rho )),
\end{equation}
where $D(\rho)$ is expressed as [8] 
\begin{multline}
D(\rho)= A_x A_y \sum_{k=0}^{\infty}\sum_{n=0}^{\infty} \tilde{B}_x(n) \tilde{B}_y(k) n!k! \sum_{i=0}^{n}\sum_{j=0}^{k} \\
\frac {(i+j)}{i!j!} \frac {a_x^i+a_y^j}{\rho^(i+j)}e^\frac{a_x+a_y}{\rho}  \Gamma (-i-j,\frac{a_x+a_y}{\rho})) 
\end{multline}
Moreover, $D( p_{N_1} \rho)$ can be derived as $D(\rho)$. Similarly, the exact expression of $C_{x_2}$ can be derived as [8]
\begin{multline}
C_{x_2}^{exact} = \frac {1}{2ln2} A_w A_y\sum_{k=0}^{\infty}\sum_{n=0}^{\infty} \tilde{B}_w(n) \tilde{B}_y(k) n!k! \sum_{i=0}^{n}\sum_{j=0}^{k} \frac {(i+j)}{i!j!}  \\ 
 \frac {a_w^i+(a_y/p_{N_1})^j}{\rho^(i+j)}e^\frac{a_w+(a_y/p_{N_1})}{\rho} \Gamma (-i-j,\frac{a_w+(a_y/p_{N_1})}{\rho}) 
\end{multline}
Moreover, the OAM channel is performing at LOS communication for the CCU and CCU has better channel condition than CEU as well. So, the exact expression of $C_{x_3}$ by OAM can be achieved as below [9,12,16-17]
\begin{equation}
C_{x_3}^{exact}=\frac{1}{2}\log_2(1+{p_{N_2} \rho }{{\mu_k^2}} ).
\end{equation}
From Eq. (22), Eq. (24) and Eq. (25), the sum capacity can be achieved as  
\begin{equation}
C_{\text{CCU}}^{exact} = \text{E}[C_{x_1}^{exact}]+\text{E}[C_{x_3}^{exact}],
\end{equation}
\begin{equation}
C_{\text{CEU}}^{exact} = \text{E}[C_{x_2}^{exact}],
\end{equation}
\begin{equation}
C_{\text{sum}}^{exact} = C_{\text{CCU}}^{exact}+C_{\text{CEU}}^{exact},
\end{equation}
where E[.] represents the expectation operator. Moreover, $C_{\text{CCU}}^{exact}$ and $C_{\text{CEU}}^{exact}$ are the exact capacity of CCU and CEU for the proposed CNOMA-OAM scheme. 
\section{Numerical Result Analysis}
\begin{figure}[t!]
\centering
\includegraphics[width=0.4\textwidth]{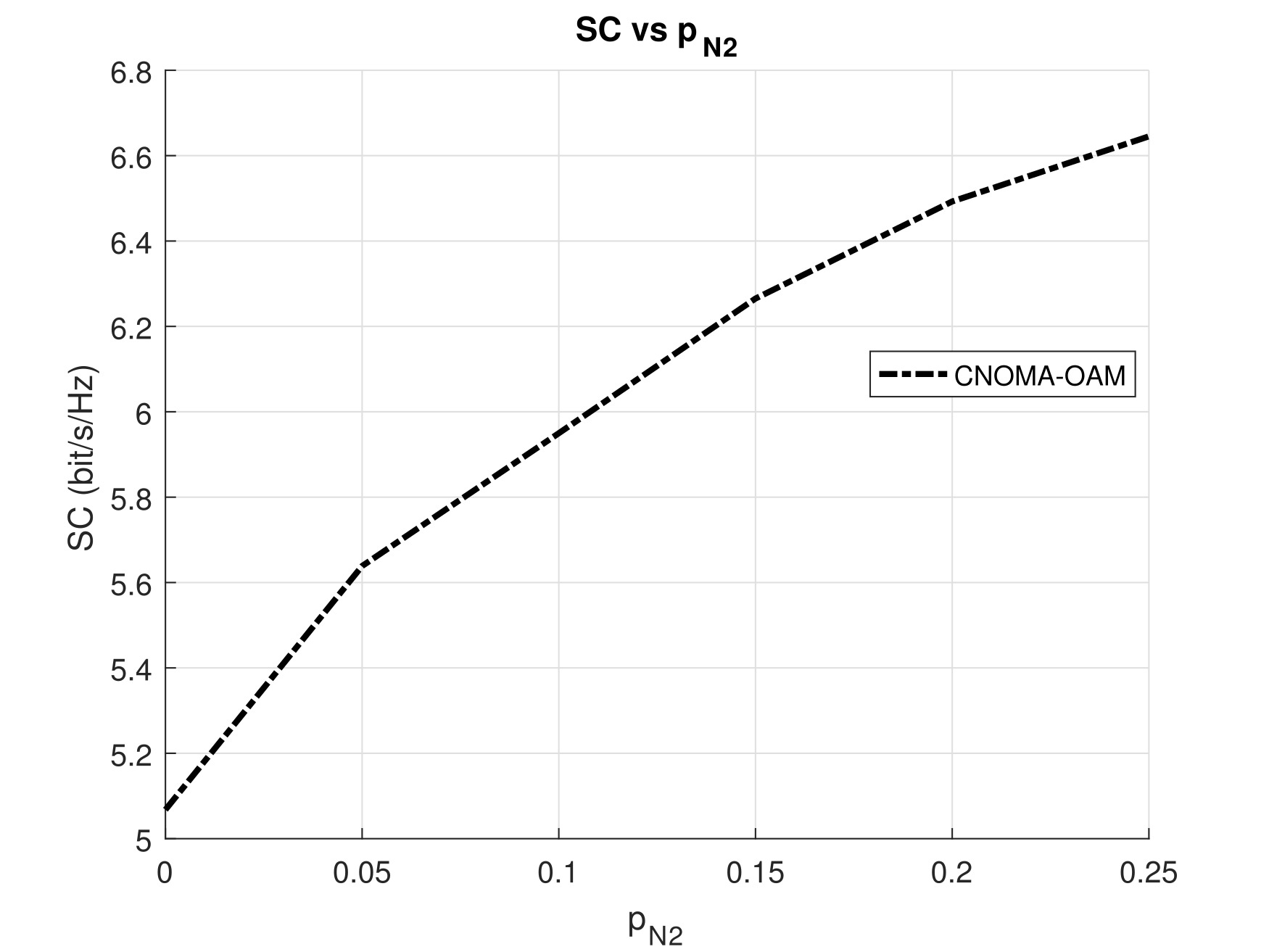}
\caption{SC comparisons  with respect to $p_{N_2}$.}
\label{image-myimage}
\end{figure}
\begin{figure}[t!]
\centering
\includegraphics[width=0.4\textwidth]{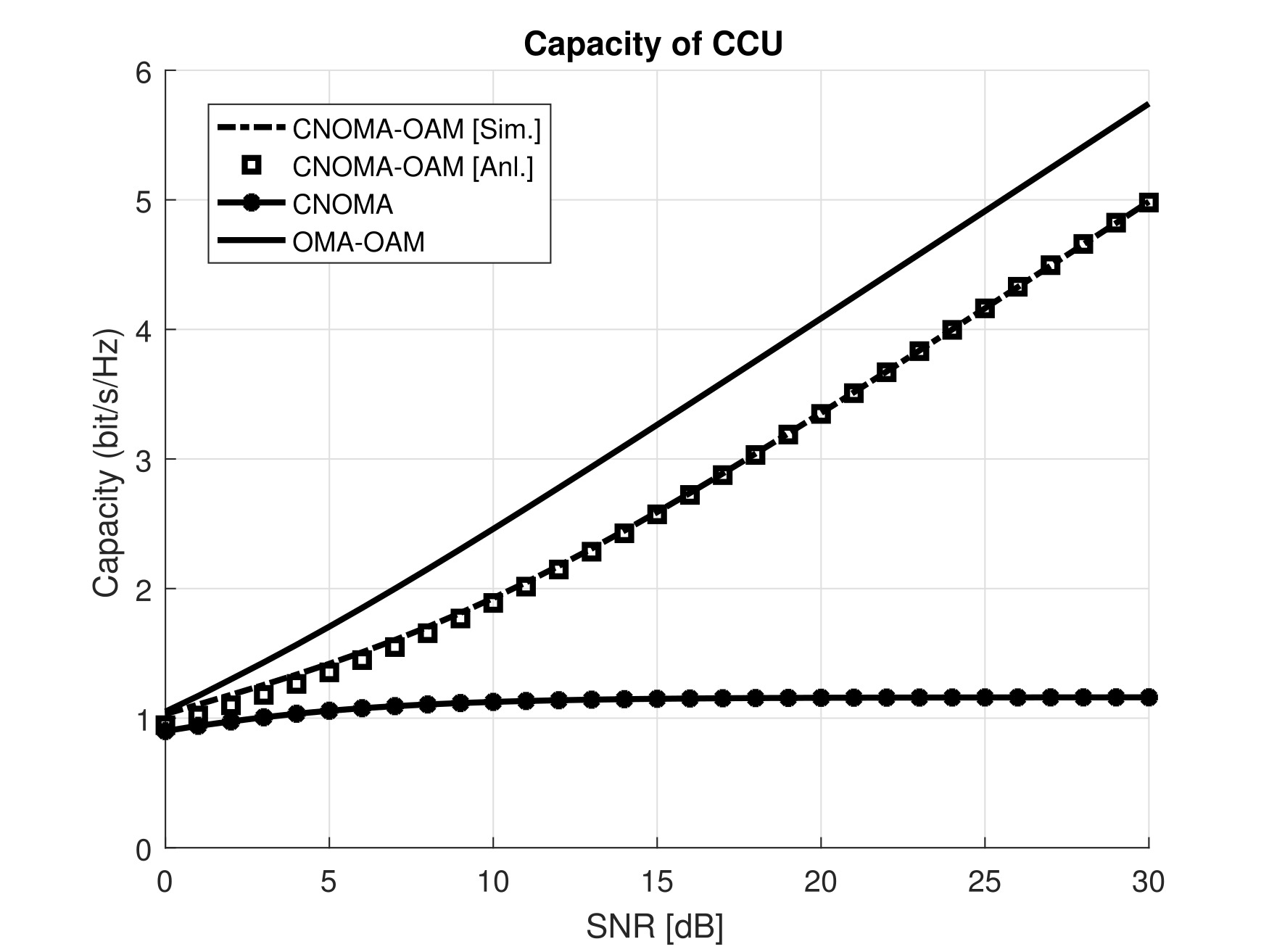}
\caption{Capacity comparisons of CCU with respect to SNR.}
\label{image-myimage}
\end{figure}
\begin{figure}[t!]
\centering
\includegraphics[width=0.4\textwidth]{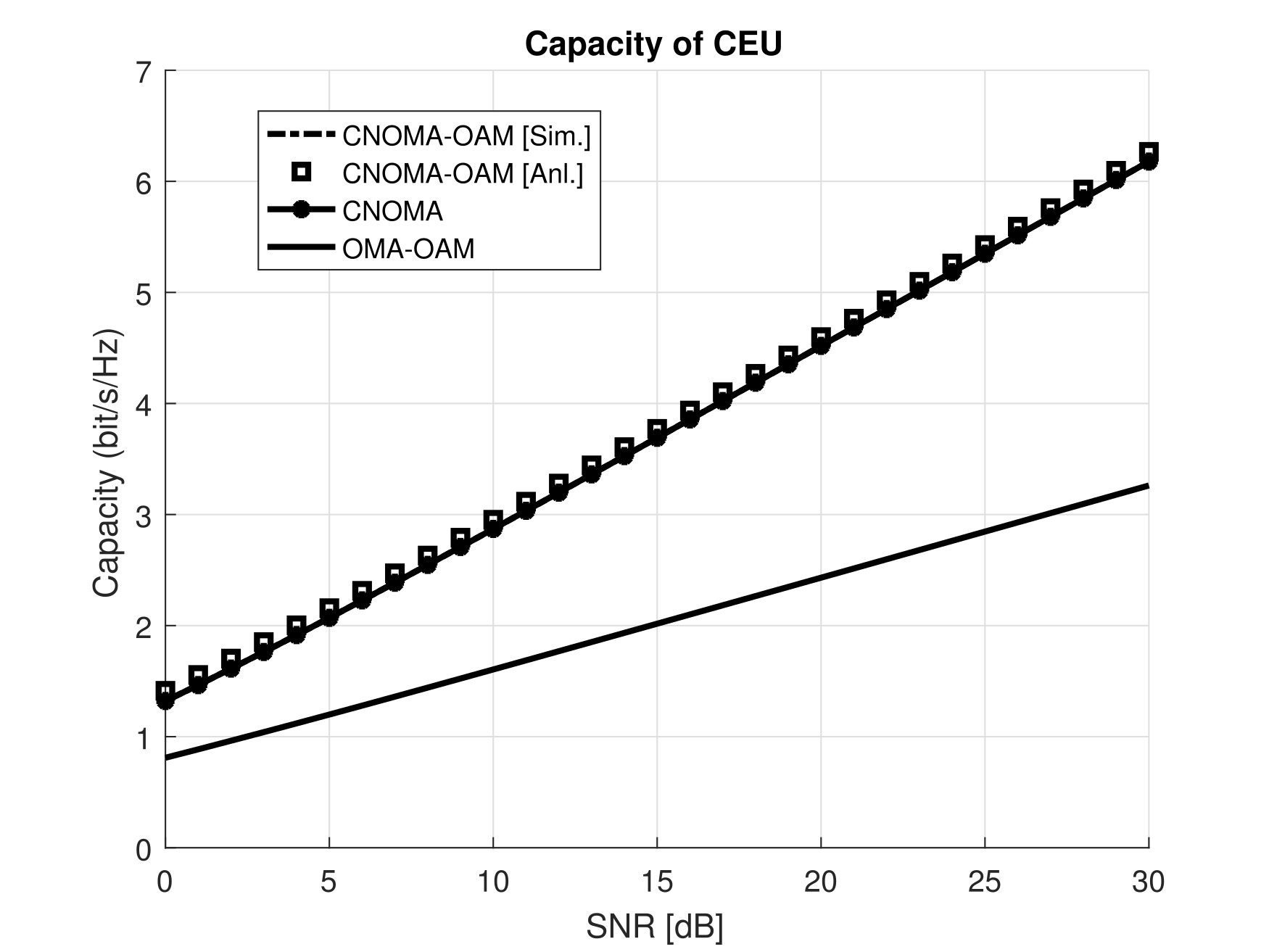}
\caption{Capacity comparisons of CEU with respect to SNR.}
\label{image-myimage}
\end{figure}
\begin{figure}[t!]
\centering
\includegraphics[width=0.4\textwidth]{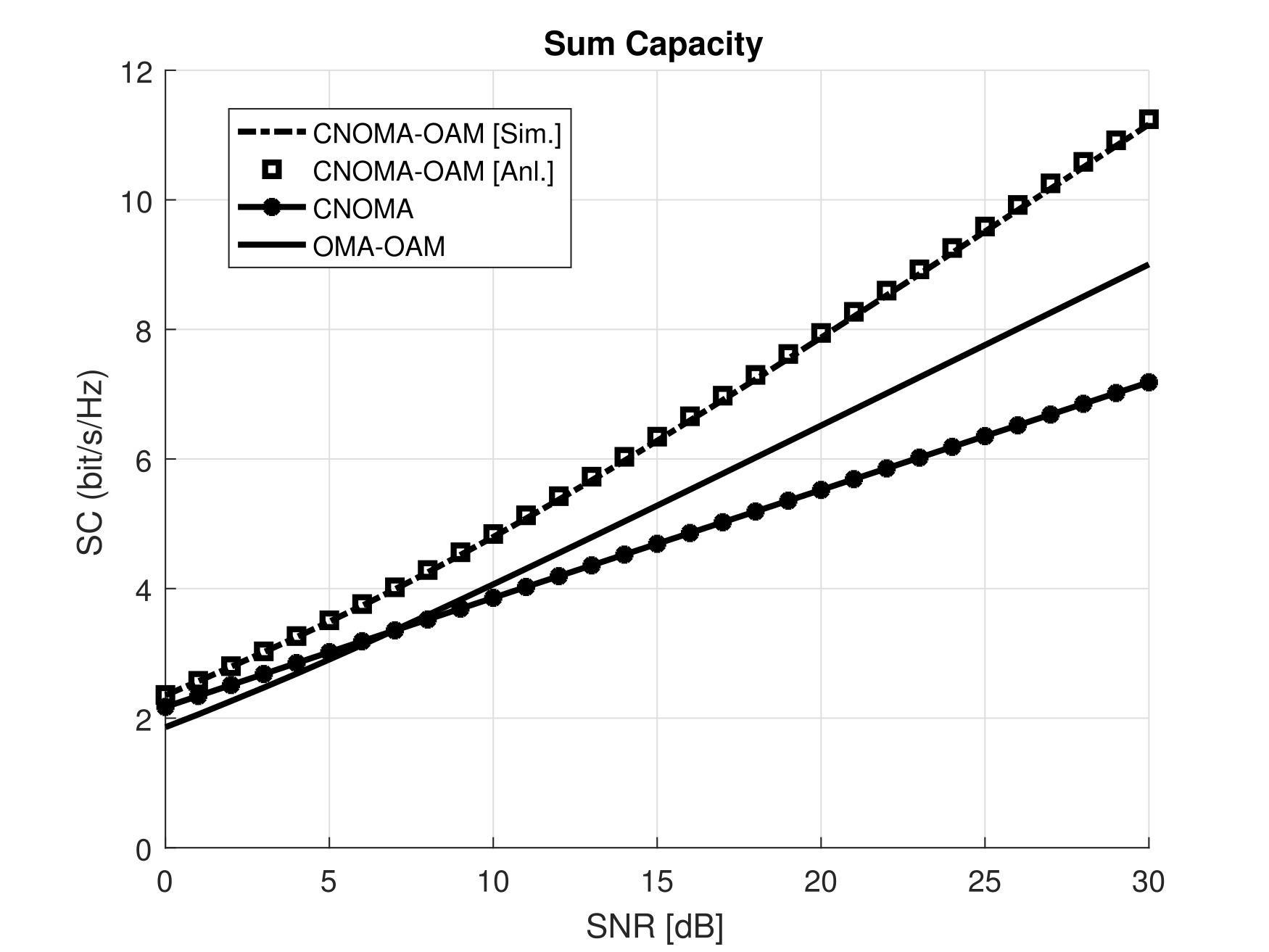}
\caption{SC comparisons with respect to SNR.}
\label{image-myimage}
\end{figure}
The Rician fading parameters are considered as $K_{SR} = K_{RD} = 5$, and $K_{SD} = 2$ [8]. 
 In addition, $d_1=0.5$ and $d_2=1$ are considered for all simulation result analysis. 
 $\Omega_{CEU}=9< \Omega_{CCU}=\Omega_{Relay}=36$ are considered for simulation [8]. Furthermore, $P=1$ is assumed here for simulation purpose. 
 \par
 The impact of the allocated power of the OAM beam over the SC for CNOMA-OAM is analyzed in Fig.3. $ \rho = 15 dB$ and $p_F=0.6$ are considered for SC versus 
$p_{N_2}$ comparison for the proposed scheme. As presented in the Fig. 3, 
$p_{N_2} = 0.2$ provides optimum SC for the proposed CNOMA-OAM scheme. Hence, 
$p_{N_1} = 0.2$, $p_{N_2} = 0.2$, and $p_F = 0.6$ are considered for capacity vs SNR comparison.       
  \par
 The capacity comparisons of CCU and CEU with respect to SNR is shown in Fig.4 and Fig.5 for the proposed and other conventional schemes. Moreover, SC comparison of CNOMA-OAM and other schemes are demonstrated in Fig. 6. CNOMA-OAM scheme provides higher SC than conventional CNOMA and OMA-OAM which is shown in Fig.6. Whereas, SC is the addition of capacity of CCU and CEU. So, the capacity enhancement of CCU in case of proposed scheme can be achieved by utilizing an additional OAM channel to transmit $x_3$ from BS to CCU simultaneously in the first time slot which is shown in Fig.4. However, OMA-OAM provides slightly higher capacity at CCU than CNOMA-OAM due to full power transmission. The achieved capacity of CEU is same for CNOMA-OAM and conventional CNOMA which is shown in Fig. 5. Moreover, CNOMA-OAM provides significantly higher capacity than OMA-OAM which illustrates in Fig. 5 as well. As a result, the SC is enhanced for the CNOMA-OAM scheme by utilizing the CNOMA and OAM concept effectively. The analytical results of the proposed CNOMA-OAM scheme are also validated by the Monte-Carlo simulation results.
\section{Conclusion}
In this letter, CNOMA-OAM is proposed to enhance SC.The numerical result analysis showed that the proposed CNOMA-OAM provides a significantly higher SC than other schemes over the Rician fading channel.
The provided analytical results are justified by the simulation results. The optimum transmitted power for OAM beam in proposed CNOMA-OAM is also analyzed. For future work, CNOMA-OAM-based simultaneous wireless information and power transfer can be studied. 

\end{document}